\documentclass[aps,prb,twocolumn,showpacs,superscriptaddress]{revtex4-1}
\usepackage{amsfonts,amsmath,amssymb}
\usepackage[colorlinks=true,linkcolor=blue,pagecolor=blue,filecolor=blue,menucolor=blue,urlcolor=blue,citecolor=blue,anchorcolor=blue]{hyperref}%
\usepackage[dvips]{graphics}
\usepackage{graphicx}
\usepackage{bm}
\usepackage{color}

\begin{document}

\title{Magnetoelectrics in Disordered Topological Insulator Josephson Junctions}
%\title{paper}
\author{I. V. Bobkova}
\affiliation{Institute of Solid State Physics, Chernogolovka, Moscow
  reg., 142432 Russia}
\affiliation{Moscow Institute of Physics and Technology, Dolgoprudny, 141700 Russia}
\author{A. M. Bobkov}
\affiliation{Institute of Solid State Physics, Chernogolovka, Moscow reg., 142432 Russia}
\author{Alexander A. Zyuzin}
\affiliation{A. F. Ioffe Physical - Technical Institute, 194021 St. Petersburg, Russia}
\author{Mohammad Alidoust }
\affiliation{Department of Physics, Faculty of Sciences, University of Isfahan, Hezar Jerib Avenue, Isfahan 81746-73441, Iran}

\date{\today}

\begin{abstract}
We theoretically study the coupling of electric charge and spin polarization in
an equilibrium and nonequilibrium electric transport across a two dimensional Josephson configuration comprised of disordered
surface channels of a three dimensional topological
insulator. In the equilibriun state of the system we
  predict the Edelstein effect, which is much
  more pronounced than its counterpart in conventional spin orbit coupled materials. Employing a quasiclassical Keldysh technique, we
demonstrate that the ground state of system can be experimentally
shifted into arbitrary macroscopic superconducting phase differences other than
the standard `$0$' or
`$\pi$', constituting a $\phi_0$-junction, solely by modulating a
quasiparticle flow injection into the junction. 
We propose a feasible
experiment where the quasiparticles are injected
into the topological insulator surface by means of a normal electrode
and voltage gradient so that oppositely oriented
stationary spin densities can be developed along the interfaces and allow for directly making use of
the spin-momentum locking nature of Dirac fermions in the surface
channels. The $\phi_0$-state is proportional to the voltage
difference applied between the injector electrode and superconducting
terminals that calibrates the injection rate of particles and,
therefore, the $\phi_0$ shift.
\end{abstract}

\maketitle

\section{introduction}

Topological state of matter has been a striking topic
during the past decade and has attracted extensive attention
both theoretically and experimentally \cite{rev1,rev2}. 
The study of nontrivial electronic structure topology goes back to the prediction of the time-reversal symmetry protected electronic states 
bound to the contact of two semiconductors with mutually inverted bands \cite{pankratov} and exotic particles in liquid $\mathrm{He}$ \cite{volovik}, which now offer promising prospects to
practical platforms such as spintronics and quantum computation \cite{rev1,rev2}. 
The extensive and growing research efforts in this context have led to the exploration of
topological insulators (TIs) \cite{ti1,ti2,ti3,ti4,ti5}, for
instance. Surface channels of a three dimensional topological insulator are conductors \cite{pankratov} while bulk material itself is insulator with a gap in its band structure.
In contrast to numerous theoretical works so far \cite{rev1,rev2},
these surface channels may not be fully ballistic and can host
unavoidable defects or other quasiparticle scattering
resources such as impurities \cite{burkov,Schwab,zyuzin}. The recent
experiments have also revealed this fact that the conduction of the
surface states are limited with a finite resistance
\cite{nowack,spanton}. This key observation clearly proves the
fundamental importance of considering the contributions and influences
of disordered motions of moving particles in the surface channels \cite{nowack,spanton,burkov,Schwab,zyuzin}. 

The surface channels may be employed as an appealing opportunity that
can support spin-momentum locked modes, namely, the orientation of
particle's spin is locked to its momentum direction
\cite{rev1,rev2}. The motion of low energy
quasiparticles in these surface states are governed by a Dirac
equation, meaning the particles' velocity is energy independent in low
energy bands. The interplay of the Dirac particles with superconductivity can result in rich and intriguing physics such as particle to antiparticle conversion \cite{rel1,rel2}. While the surface states themselves may not be superconducting, they may however adopt superconducting properties by proximity to a
superconducting electrode \cite{exp_ti1,exp_ti2,exp_ti3,exp_ti4,exp_ti5,exp_ti6}. Therefore, such a platform can provide the unique possibility
to materialize the interplay of relativity and superconductivity
in laboratory and possibly make use of its advantages in functional devices and technological ways \cite{phi0}.

Motivated by the recent experimental progresses in fabricating
TI-superconductor (S) heterostructures
\cite{exp_ti1,exp_ti2,exp_ti3,exp_ti4,exp_ti5,exp_ti6}, we here study one of the
manifestations of interplay between the superconductivity and
spin-momentum locking nature of Dirac fermions on the surface of a 3D
TI: The coupling between spin and charge degrees of freedom in a two
dimensional Josephson junction made of disordered surface states of a
three dimensional TI under equilibrium and nonequilibrium conditions. 

In the equilibrium state, we predict that the system
responses to a dc supercurrent flowing across the junction by
developing a stationary spin density oriented along the junction
interfaces. %in response to a dc Josephson current flowing across the
            %junction. 
This phenomenon can be viewed as a direct magnetoelectric effect i.e.
the Edelstein effect. 
In the {\it conventional} intrinsic spin orbit coupled metals, analogous
effect was first theoretically predicted in
Refs. \onlinecite{aronov89,edelstein90} and later observed
experimentally in Refs. \onlinecite{kato04,silov04}. 
It was shown that spin
polarization can be produced in the spin orbit coupled metals by externally applied electric field.
\cite{aronov89,edelstein90,kato04,silov04}. The magnetoelectric
polarizability was also discussed in the normal phase of topological
insulators \cite{rev1,rev2,essin09}.  
We note that similar magnetoelectric effect was also predicted for
bulk superconductors and superconducting heterostructures by means of
supercurrent due to the
spin orbit interactions
\cite{edelstein95,edelstein05,malshukov08}. Here we develop a theory
to this phenomenon in S-TI-S heterostructures and show that a
topological insulator can support much more pronounced signatures than
those of conventional spin orbit coupled metals.

We also demonstrate an inverse coupling between the charge and spin
degrees of freedom in S-TI-S heterostructures. In particular, it is
shown that the injection of quasiparticle current at the
middle of junction in opposite directions yields spontaneous
magnetizations with opposite orientations in each segment. The
current-induced magnetization is directly coupled to the phase
difference between the superconducting terminals. This phenomenon
renders the minimum of junction free energy into a superconducting
phase difference between the S terminals
$\phi = \phi_0$, which is generally unequal to $0$ or $\pi$. The $\phi_0$ is proportional
to the voltage difference between the injector electrode and
superconducting terminals and inverse of Fermi velocity on the surface
states. $\phi_0$-junctions were also predicted in 
ferromagnetic Josephson junctions with spin orbit interaction
\cite{krive04,braude07,reinoso08,buzdin08,kuzmanovski16} and may be viewed as an
inverse magnetoelectronic effect in such hybrids
\cite{konschelle15}. 

Nearly all of the past theoretical works on the spin orbit
coupled Josephson structures involves ferromagnetism or an external
magnetic field as key ingredients to establish the magnetoelectric
effect. In this paper, however, we demonstrate that an internal
Zeeman-like term can be generated and electrically controlled in
S-TI-S heterostructures (in the absence of any ferromagnetic elements
or an external magnetic field) simply by means of quasiparticles
injection. The Zeeman-like term causes an anomalous phase shift
$\phi_0$ in the supercurrent that can be electrically controlled via
the quasiparticle flow injection generated by a voltage difference. 

So far, the electric control of the superconducting
  critical temperature \cite{bobkova11,bobkova15,ouassou16}, the
  magnitude of the Josephson current \cite{heikkila00,bobkova12},
  switching between $0$ and $\pi$ states \cite{volkov95,wilhelm98,baselmans99,huang02,crosser08,bobkova10}, and spontaneously
accumulated spin currents \cite{ald15} have been discussed in hybrid
structures with spin orbit coupling. The electric
control of the anomalous phase shift was experimentally realized 
in superconducting heterostructures on the basis of a quantum
wire\cite{phi0}. This kind of controll also was theoretically proposed in
S/silicene/S heterostructures \cite{kuzmanovski16}. Nonetheless, in
these proposals, the anomalous phase shift occurs only in the presence
of an externally applied magnetic field or ferromagnetic elements. Our findings offer the ability of
integrating the superconducting topological insulator nanostructures
into electronic circuits without the requirement to apply magnetic
field or making use of magnetic elements.

The paper is organized as follows. In Sec.~\ref{method}, we summarize
the theoretical framework used and basic assumptions made for studying
a disordered topological insulator Josephson junction under
nonequilibrium conditions. The direct magnetoelectric effect in S-TI-S
heterostructures is presented in Sec.~\ref{direct}. The electrically
controllable inverse magnetoelectric effect is discussed in
Sec.~\ref{inverse}. We finally summarize concluding remarks in Sec.~\ref{conclusions}.

\section{nonequilibrium Keldysh  technique}\label{method}

\begin{figure}[!tbh]
  %\centerline{\includegraphics[clip=true,width=3.5in]{fig4.eps}}
           %\centerline{\includegraphics[clip=true,width=2.5in]{fig1b.eps}}
  \begin{minipage}[b]{\linewidth}
     \centerline{\includegraphics[clip=true,width=2.8in]{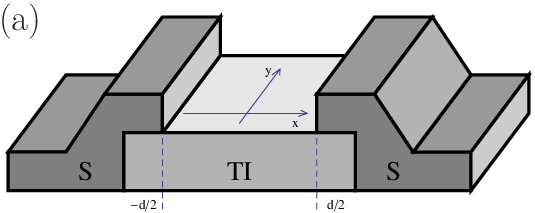}}
     \end{minipage}\hfill
    \begin{minipage}[b]{\linewidth}
   \centerline{\includegraphics[clip=true,width=2.2in]{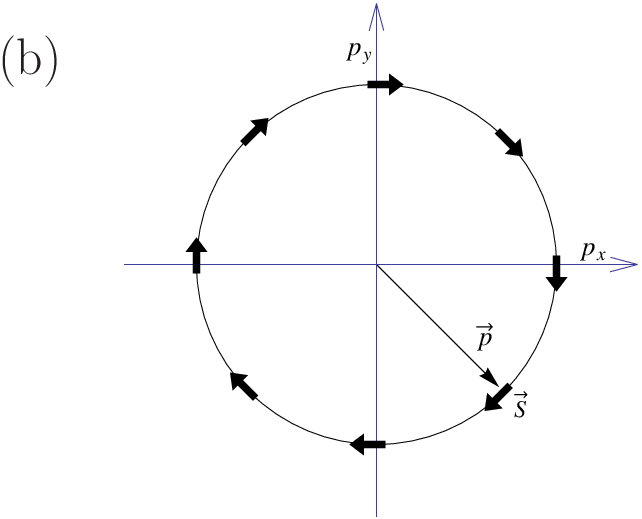}}
  \end{minipage}
   \caption{(a) Schematic of the two dimensional S-TI-S hybrid
     junction. The surface
     states of a three dimensional topological insulator is sandwiched
     between two $s$ wave superconductors. The junction length along
     the $x$ direction is equal to $d$ while its width along $y$ is assumed
     enough large so that the influences of boundaries in this direction are
     negligible. The junction plane resides
     in the $xy$ plane and S-TI interfaces are located at $x=\pm d/2$. (b) The spin-momentum
     locking phenomenon in the conduction band of surface channels.}   
\label{system}
\end{figure}

Numerous systems in laboratory deal with
nonequilibrium processes. To theoretically describe a system that
experiences nonequilibrium situations, one powerful approach is the Keldysh technique within the
Green function framework \cite{keldysh,mahan}.  
In this section, we start with the Hamiltonian of surface states of a
three dimensional topological insulator and generalize the Keldysh
technique to a Josephson structure made of the disordered surface
states of a three dimensional TI under nonequilibrium.
The system under consideration is schematically shown in Fig.~\ref{system}(a). 

The Hamiltonian that describes the Rashba type surface states in the
presence of an in-plane exchange field $\bm h(\bm r)=(h_x(\bm r), h_y(\bm r), 0)$ reads: 
\begin{equation}
\hat H(\bm r)=-i\alpha (\bm \nabla \times \bm e_z)\hat {\bm \sigma}+\bm h(\bm r)\hat {\bm \sigma} + V_{imp}(\bm r)-\mu
\label{h},
\end{equation}
\begin{equation}
\hat H=\int d^2 \bm r' \hat \Psi^\dagger (\bm r')\hat H(\bm r')\hat \Psi(\bm r')
\label{H},
\end{equation}
in which $\hat \Psi=(\Psi_\uparrow, \Psi_\downarrow)^T$, $\alpha$ is
the Fermi velocity, $\bm e_z$ is a unit vector normal to the surface
of TI (see Fig. \ref{system}(a)), $\mu$ is the chemical potential, and
$\hat {\bm \sigma}=(\sigma_x, \sigma_y, \sigma_z)$ is a vector of
Pauli matrices in the spin space. Note that the Dirac type of
Hamiltonian would only change the notations \cite{zyuzin}. We assume
that the system involves nonmagnetic impurities that can be described
by a Gaussian scattering potential: $V_{imp}(\bm r)=\sum \limits_{\bm r_i} V_i \delta(\bm r - \bm r_i)$
\begin{eqnarray}
\big\langle V(\bm r)V(\bm r') \big\rangle =  \frac{1}{\pi \nu \tau}\delta(\bm r-\bm r'),~~ \nu=\frac{\mu}{2\pi \alpha^2}
\label{impurities}.
\end{eqnarray}
Here $\tau$ is the mean free time of quasiparticles and $\nu$ is the density of states at the Fermi level of the normal state of TI.

The advanced ($A$), retarded ($R$), and Keldysh ($K$) blocks of Gor'kov Green function
in the Keldysh technique are defined as follows:
\begin{subequations}
\begin{equation}
G_{\alpha \beta}^{R,A}(\bm r_1, \bm r_2, t_1, t_2)=\mp i \Theta_{t_{1,2}}
\big\langle \big\{ \Psi_\alpha (\bm r_1,t_1),\Psi_\beta^\dagger (\bm r_2,t_2) \big\} \big\rangle 
\label{GRA},
\end{equation}
\begin{equation}
F_{\alpha \beta}^{R,A}(\bm r_1, \bm r_2, t_1, t_2)=\pm i \Theta_{t_{1,2}}
\big\langle \big\{ \Psi_\alpha (\bm r_1,t_1),\Psi_\beta (\bm r_2,t_2) \big\} \big\rangle 
\label{FRA},
\end{equation}
\begin{equation}
F_{\alpha \beta}^{\dagger R,A}(\bm r_1, \bm r_2, t_1, t_2)=\pm i \Theta_{t_{1,2}}\big\langle \big\{ \Psi_\alpha^\dagger (\bm r_1,t_1),\Psi_\beta^\dagger (\bm r_2,t_2) \big\} \big\rangle 
\label{F+RA},
\end{equation}
\begin{equation}
\tilde G_{\alpha \beta}^{R,A}(\bm r_1, \bm r_2, t_1, t_2)=\mp i \Theta_{t_{1,2}}
\big\langle \big\{ \Psi_\alpha^\dagger (\bm r_1,t_1),\Psi_\beta (\bm r_2,t_2) \big\} \big\rangle 
\label{tildeGRA},
\end{equation}
\begin{equation}
G_{\alpha \beta}^K(\bm r_1, \bm r_2, t_1, t_2)=- i \big\langle \big[ \Psi_\alpha (\bm r_1,t_1), \Psi_\beta^\dagger (\bm r_2,t_2) \big] \big\rangle,
\label{GK}
\end{equation}
\end{subequations}
where $\{...,...\}$ and  $[...,...]$ stand for anticommutator and
commutator relations, respectively. Also, the time ordering operator
is shown by a step function in time $\Theta_{t_{1,2}}\equiv \Theta({t_{1}-t_2)}$. The other
components of Keldysh block i.e. $F_{\alpha \beta}^K$, $F_{\alpha
  \beta}^{\dagger K}$, and $\tilde G_{\alpha \beta}^K$ are related to
this component
Eq.~(\ref{GK}) the same way as those of the retarded and advanced
components given above, namely Eqs.~(\ref{FRA})-(\ref{tildeGRA}), to Eq.~(\ref{GRA}). 

We now introduce the following matrix Green function in the Nambu space:
\begin{equation}
\check G^R=\left(
\begin{array}{cc} 
\hat G^R_{new} & \hat F^R_{new}  \\
 \hat F^{\dagger R}_{new} &  \hat {\tilde G}^R_{new} \\ 
\end{array}
\right)=\left(
\begin{array}{cc} 
\hat G^R_{old} & \hat F^R_{old} i \sigma_y \\
i\sigma_y \hat F^{\dagger R}_{old} & -\sigma_y \hat {\tilde G}^R_{old} \sigma_y \\ 
\end{array}
\right)
\label{matrix_ph},
\end{equation}
where $\hat G^R_{old} (\hat F^R_{old}, \hat F^{\dagger R}_{old}, \hat
{\tilde G}^R_{old})$ are $2 \times 2$ matrices in the spin space. The
elements of these matrices are given above by
Eqs.~(\ref{GRA})-(\ref{tildeGRA}). This definition can be extended to
the advanced and keldysh blocks $\check G^A$ and $\check G^K$. 
It is also convenient to introduce a full $8 \times 8$ Green function
in the Keldysh space as follows: 
\begin{equation}
\check G=\left(
\begin{array}{cc} 
\check G^R & \check G^K \\
0 & \check G^A \\ 
\end{array}
\right)
\label{matrix_keldysh}.
\end{equation}
Therefore, by averaging the Green function over the impurity scattering potential in the Born approximation we find the following Gor'kov equation:
\begin{eqnarray}
\left(
\begin{array}{cc} 
i{\partial_{t_1}}-\hat H(\bm r_1) & 0 \\
0 & -i{\partial_{t_1}}-\sigma_y\hat H^*(\bm r_1)\sigma_y \\ 
\end{array}
\right)\check G= \nonumber \\
\delta(\bm r_1-\bm r_2)\delta(t_1-t_2)+\frac{1}{\pi \nu \tau}\check G(\bm r_1,\bm r_1)\check G(\bm r_1,\bm r_2)
\label{Gor'kov_imp}.
\end{eqnarray} 
Here, the two time dependent products of $AB$ operators is equivalent to $AB(t_1,t_2) \equiv \int dt' A(t_1,t')B(t',t_2)$. 

We consider a situation where the chemical potential $\mu$ is the
largest energy scale in the system and hence a quasiclassical
approximation is the well suited framework to describe the system. We
now can introduce the quasiclassical Green function
\begin{equation}
\check g =\frac{i}{\pi} \int d \xi \check G =
\left(
\begin{array}{cc}
\hat g & \hat f \\
\hat {\tilde f} & \hat {\tilde g} \\
\end{array}
\right),
\label{GF_quasiclassical}
\end{equation}
where a Fourier transformation of the Green function with respect to
the relative space arguments and an integration over $\xi=\alpha p -
\mu$ are performed. Furthermore, following the standard procedures \cite{zyuzin}, we find the Eilenberger equation for $\check g$:
\begin{eqnarray}
&&\frac{\alpha}{2}\left\{ \hat {\bm \eta}, \bm \nabla \check g \right\} = \nonumber \\&&-\hat \tau_z \partial_{t_1} \check g - \partial_{t_2} \check g \hat \tau_z + 
\left[ \check g, i \bm h (\bm R) \hat {\bm \sigma} \hat \tau_z + i\mu \hat {\bm \eta} \bm n_{F} + \frac{\big\langle \check g \big\rangle}{\tau} \right] 
\label{eilenberger},~~~~
\end{eqnarray}
in which $\hat \eta = (- \sigma_y, \sigma_x ,0)$, $\bm n_{F}=\bm p_F/|\bm
p_F|$, and $\tau_i$, $i=x,y,z$ are Pauli matrices in the particle-hole
space. This is a nonequilibrium generalization to the Eilenberger
equation derived in Ref.~\onlinecite{zyuzin} for the equilibrium
situations. Recently, the analogous quasiclassical equation was also discussed for Dirac edge and surface electrons with another type of spin-momentum locking ${\bm p}_F \bm \sigma$ \cite{hugdal16}. In the stationary situations, that we consider throughout the
paper, one can perform a Fourier transformation with respect to
$t_1-t_2 \to \varepsilon$. Thus, the Eilenberger equation can be expressed by:
\begin{equation}
\frac{\alpha}{2}\left\{ \hat {\bm \eta}, \bm \nabla \check g \right\} = \left[ \check g, -i\varepsilon \hat \tau_z+i \bm h (\bm R) \hat {\bm \sigma} \hat \tau_z + i\mu \hat {\bm \eta} \bm n_{F} + \frac{\big\langle \check g \big\rangle}{\tau} \right] 
\label{eilenberger_stat}.
\end{equation}
To find a general solution to Eq.~(\ref{eilenberger_stat}), the Green
function can be expanded by the Pauli matrices as follows; 
\begin{equation}
\check g (\varepsilon, \bm R, \bm n_{F})=\frac{1}{2}\hat g'+\frac{1}{2}\hat g''\hat {\bm \eta} \bm n_{F}+\hat g_{||}\bm n_{F}\hat{\bm \sigma} + \hat g_z \sigma_z
\label{Pauli_expansion},
\end{equation}
where $\hat g'$,$\hat g''$, $\hat g_{||}$, and $\hat g_z$ depend on
$(\varepsilon, \bm R, \bm n_{F})$ and are $4\times 4$ matrices in the
particle-hole and Keldysh space. It can be shown that $\hat g_{||}$
and $\hat g_z$ are smaller than both $\hat g'$ and $\hat g''$ by a
factor of order of $(\tau^{-1},h,\varepsilon)/\mu \ll 1$. Therefore,
one can safely neglect these terms in the expansion without missing
significant information. The Eilenberger equation in this
approximation decomposes into two coupled equations for $\hat g'$ and
$\hat g''$:
\begin{subequations}
\begin{equation}
\alpha \bm n_{F} \bm \nabla \hat g''  = \left[\hat g', -i\varepsilon
  \hat \tau_z +\frac{\big\langle \check g' \big\rangle}{2\tau} \right] +
\left[\hat g'',i h_1 \hat \tau_z + \frac{\big\langle \hat g''
    \big\rangle}{2\tau}\right],~~ 
\label{g'_g''1}
\end{equation}
\begin{equation}
\alpha \bm n_{F} \bm \nabla \hat g'  = \left[\hat g'', -i\varepsilon \hat \tau_z +\frac{\big\langle \hat g' \big\rangle}{2\tau}\right] + \left[\hat g', i h_1 \hat \tau_z + \frac{\big\langle \hat g'' \big\rangle}{2\tau}\right],~~
\label{g'_g''2}
\end{equation}
\end{subequations}
where $h_1=h_x n_{F,y}-h_y n_{F,x}$. It is apparent that Eqs.~(\ref{g'_g''1})
and (\ref{g'_g''2}) allow for a solution of $\hat g'=\hat g''$. This
means that the spin structure of full Green function is determined by
$(1+\hat {\bm \eta} \bm n_{F})/2$. Physically, this operator projects the
Green function onto the conduction band of TI surface states as
schematically depicted in Fig.~\ref{system}(b). We assume that this
band can describe the surface electrons of TI and take
the spinless Green function $\hat g'$ in the rest of our
calculations. Now, equations (\ref{g'_g''1}) and (\ref{g'_g''2})
reduce to a single spinless equation so that $\hat g'^2 $ satisfies an
equation of the same form. Also, the normalization condition
i.e. $\hat g'^2 =1$ is a solution to this equation.

In the diffusive regime, where the quasiparticles' motion is fully
randomized by strong scattering resources so that $\varepsilon, h
<\tau^{-1}$, the Eilenberger equation reduces to the Usadel
equation\cite{usadel}. To have a self-contained presentation, we
here give a short recap of basic notations and equations generalized
for a nonequilibrium situation \cite{zyuzin}.
In the diffusive regime, the Green function can be expanded through the first two harmonics:
\begin{equation}
\hat g' (\varepsilon, \bm R, \bm n_{F})=\hat g_s(\varepsilon, \bm R)+\bm n_{F}\hat {\bm g}_a(\varepsilon, \bm R)
\label{g1}, 
\end{equation}
where the zeroth harmonic (isotropic) is larger than the first harmonic: $\hat g_s \gg \hat {\bm g}_a$.
Following the derivation steps described in Ref.~\onlinecite{zyuzin},
one obtains the Usadel equation to $\hat g_s$: 
\begin{equation}
D \hat {\bm \nabla} (\hat g_s \hat {\bm \nabla} \hat g_s)=\left[-i\varepsilon \hat \tau_z, \hat g_s\right]
\label{usadel_stat},
\end{equation}
where $D=\alpha^2 \tau$ is the diffusion constant and the first harmonic term can be expressed in terms of $\hat g_s$ as follows: 
\begin{equation}
\hat {\bm g_a}=-2\alpha \tau (\hat g_s \hat {\bm \nabla} \hat g_s)
\label{g_antisymmetric}.
\end{equation}
The operator $\hat {\bm \nabla}$ is defined for the Rashba type band
structure of surface states as follows:
\begin{equation}
\hat {\bm \nabla} X = \bm \nabla X +\frac{i}{\alpha}\left( h_x {\bm e}_y - h_y {\bm e}_x \right)\left[ \tau_z, X \right]
\label{nabla_long}.
\end{equation}
Note that the formalism can be straightforwardly extended to the
Dresselhaus type \cite{zyuzin}.
In the junction configuration systems, the Usadel equation should be supplied
by appropriate boundary conditions. Here, we consider low transparent
tunnelling processes at the TI-S interfaces shown in
Figs. \ref{system}(a) and \ref{sketch}. This boundary condition is
experimentally relevant and was discussed in
Refs.~\onlinecite{bc1,bc2} for conventional metallic junctions. We assume that the
$s$ wave superconducting terminals with a gap of $\Delta$ in their
energy spectrums are in the equilibrium state and thus can be described by
the bulk solution $\hat g_{SC}$ at the boundaries \cite{zyuzin}:
\begin{equation}
2 \gamma \hat g_s \bm n \hat {\bm \nabla}\hat g_s=\left[ \hat g_s, \hat g_{SC} \right]
\label{boundary_condition},
\end{equation}
where $\gamma$ is the ratio of resistance of the interface barrier per
unit area to the resistivity of the TI surface states and $\bm n$ is
the unit vector normal to the interface (it points towards the
superconductor region). In the tunneling low proximity regime we consider throughout
our calculations, the opacity of interfaces $\gamma$ should be
sufficiently large to satisfy this condition. The superconducting bulk
solution can be given by:
\begin{eqnarray}
\hat g_{SC}^{R,A}=\pm \frac{{\rm sgn}\varepsilon}{\sqrt{(\varepsilon \pm i\delta)^2-\Delta^2}}\left[ \varepsilon \tau_z + \Delta \tau_+ - \Delta^* \tau_- \right] \nonumber \\
\hat g_{SC}^K=\left( \hat g_{SC}^R-\hat g_{SC}^A \right)\tanh \frac{\varepsilon}{2T}~~~~~~
\label{G_sc_bulk},
\end{eqnarray}
where $\tau_{\pm}=(\tau_x \pm i\tau_y)/2$, $\delta$ is an
infinitesimal  positive value and the system temperature
is denoted by $T$.

In the context of quantum transport, the electric current is one of the most
important physical quantities that can explain transport experiments. The electric current density across the surface channels
of TI junction we consider can be expressed by:
\begin{equation}
\bm j =- \frac{i e \alpha}{4}\lim \limits_{\bm r \to \bm r'}{\rm Tr}_4 \int \frac{d \varepsilon}{2 \pi} \hat {\bm \eta} \hat \tau_z \check G^K (\bm r, \bm r', \varepsilon)
\label{current_Gor'kov}.
\end{equation}
Rewriting the current density via the quasiclassical Green functions in the diffusive limit we obtain:
\begin{equation}
\bm j =\frac{\sigma_N}{8e}{\rm Tr}_2 \int d \varepsilon \left[ \hat \tau_z \left( \hat g_s \hat {\bm \nabla} \hat g_s \right)^K \right] 
\label{current_Usadel},
\end{equation}
where $\sigma_N=2e^2\nu D$ is the normal state conductivity.

\section{magnetoelectric effect}

\label{direct}

In this section, we develop a theory of the dissipationless direct
magnetoelectric effect to the  disordered topological insulator
Josephson junction depicted in Fig. \ref{system}(a). As it was
mentioned earlier in the introduction, in response to a dc Josephson
current flowing across the TI junction, a stationary spin density
$S_y$ oriented along the junction interface can be generated. To
uncover this phenomenon, we first evaluate the average spin
polarization in TI which is:
\begin{equation}
\bm S =\frac{1}{2} \big\langle \hat \Psi^\dagger(\bm r,t)\hat {\bm \sigma} \hat \Psi (\bm r,t)  \big\rangle
\label{spin_general}.
\end{equation}
In terms of the Green function, the components of spin polarization take the
following form:
\begin{eqnarray}
S_i =-\frac{i}{8}{\rm Tr}_4 \int \frac{d\varepsilon}{2\pi} \frac{d^2 p}{(2 \pi)^2} \hat \sigma_i \hat \tau_z \check G^K(\bm p, \bm R, \varepsilon)= \nonumber \\ 
-\frac{\nu}{16}{\rm Tr}_4 \int d \varepsilon \hat \sigma_i \hat \tau_z \big\langle \check g^K \big\rangle 
\label{spin_quasiclassical}.~~~~~~~~~~~~
\end{eqnarray}
Within the quasiclassical diffusive regime we consider for the
disordered topological insulator, the spin polarization components can
be rewritten as 
\begin{eqnarray}
S_i =-\frac{\nu}{16}{\rm Tr}_4 \int d \varepsilon \hat \sigma_i \hat \tau_z \big\langle \frac{1+\hat {\bm \eta}\bm n_F}{2}  (\hat g_s+\hat {\bm g}_a \bm n_F)^K \big\rangle \nonumber \\
= -\frac{\nu}{64}{\rm Tr}_4 \int d \varepsilon \left[ \hat \sigma_i \hat \tau_z \hat {\bm g}_a^K \hat {\bm \eta} \right] 
\label{spin_TI}.~~~~~~~~~~~~
\end{eqnarray}
Substituting $\hat {\bm g}_a^K$ from Eq.~(\ref{g_antisymmetric}) into Eq.~(\ref{spin_TI}), we obtain: 
\begin{eqnarray}
S_{x,y} =\pm \frac{\nu}{32}{\rm Tr}_2 \int d \varepsilon \left[ \hat \tau_z \hat {\bm g}_a^K \bm e_{y,x} \right]= \nonumber \\
\mp \frac{\nu}{16}\alpha \tau {\rm Tr}_2 \int d \varepsilon \left[ \hat \tau_z \hat g_s \hat \partial_{y,x} \hat g_s \right]^K  
\label{spin_xy}.
\end{eqnarray}
Comparing the last expression to Eq.~(\ref{current_Usadel}), we see
that 
\begin{equation}
S_{x,y}=\mp \frac{1}{4e\alpha}j_{y,x}
\label{ee}.
\end{equation}
That is, the electric current and electron spin polarization orientation are
perpendicular while their amplitudes are directly proportional.

When a superconducting phase gradient $\phi$ is applied
between the superconducting terminals, a dc Josephson current flows
across the junction. In the tuneling and low proximity limit, the
supercurrent has a standard sinusoidal relation $I=I_c \sin \phi$ in
which $I_c$ is the critical supercurrent \cite{zyuzin}. Therefore,
considering the spin polarization relation given above, we see that the
spin polarization can be simply controlled by the superconducting
phase difference $\phi$.
This is a generic phenomenon and occurs in ballistic systems as
well. To derive an expression for the spin density in the ballistic
regime, we consider a surface channel with superconductivity where the
superconducting phase $\phi$ experiences a coordinate gradient so that ${\bm
  \nabla}\phi(\mathbf{r}) \neq 0$. The spin density in momentum representation and in the limit when the superconducting gap is much smaller than the temperature is given by
\begin{eqnarray}
&\bm S(\mathbf{q}) =-\frac{1}{2}\text{Tr}_2T\sum \limits_{\omega_n}\int\frac{d{\bm p}d{\bm
    q'}}{(2\pi)^4}\hat{{\bm \sigma}}\Big[ \hat{G}(i\omega_n, {\bm p}+\frac{{\bm
  q}}{2})\hat{\Delta}({\bm q'}+\frac{{\bm q}}{2}) \times\nonumber\\&
\hat{G}^\dag(-i\omega_n, -{\bm p}+{\bm q'})\hat{\Delta}^\dag({\bm q'}-\frac{{\bm q}}{2}) \hat{G}(i\omega_n, {\bm p}-\frac{{\bm q'}}{2})
\Big],
\end{eqnarray}
where $\hat{\Delta}(\mathbf{q})=i\hat{\sigma}\Delta(\mathbf{q})$.
The
Green function in the Matsubara representation reads:
\begin{eqnarray}\label{Green_Magnetoelectric}
&&{\hat{G}}(i\omega_n, \bm p)=\nonumber \\&&\left[ i\omega_n - \hat H(\bm p) \right]^{-1}= 
\frac{1}{2}\sum \limits_{{s} = \pm 1} \frac{1+{s} \left[ \bm n_F \times {\bm e}_z \right]{\hat{\bm \sigma}}}{i\omega_n+\mu-{s}\alpha p}
\label{G_normal}.
\end{eqnarray}
%
%\begin{eqnarray}
%\hat{G}({\bm p},\omega_n)=\frac{1}{2}\Big[\frac{1+{\bm \eta}{\bm p}}{i\omega_n+\mu-\alpha|{\bm p}|}+\frac{1-{\bm \eta}{\bm p}}{i\omega_n+\mu+\alpha|{\bm p}|}\Big].
%\end{eqnarray}
Following similar steps as Ref. \onlinecite{edelstein95}, one
obtains the spin density in the real space representation as:
\begin{eqnarray}
&{\bm S}(\mathbf{r})=-i \text{Tr}_2 \frac{\hat{{\bm \sigma}} (\hat{{\bm
  \eta}}\cdot{\bm F}(\mathbf{r}))}{16 \pi \alpha} T\sum\limits_{\omega_n>0}\Big[
1+\frac{\mu}{\omega_n}
\text{atan}\frac{\mu}{\omega_n}+\nonumber\\ &
\frac{\omega_n}{2 \mu}\Big\{ \frac{\pi}{2} -\text{atan}[\frac{\mu}{2\omega_n}(1-\frac{\omega_n^2}{\mu^2})]\Big\} \Big]\frac{1}{\omega_n^2},
\end{eqnarray}
where
$
{\bm F}(\mathbf{r})=({\bm \nabla}\Delta^*(\mathbf{r}))\Delta(\mathbf{r})-\Delta^*(\mathbf{r})({\bm \nabla}\Delta(\mathbf{r})).\nonumber
$
Considering $\Delta(\mathbf{r}) = |\Delta| e^{i\phi (\mathbf{r})}$ one finds that
at the charge neutrality point at which $\mu=0$ the spin density reduces to:
\begin{equation}
S_{x,y}(\mathbf{r}) = \mp \frac{|\Delta|}{16\pi \alpha}\frac{|\Delta|}{T}\partial_{y,x}\phi(\mathbf{r}),
\end{equation}
while within $\mu>> T_c$ regime one arrives at:
\begin{equation}
S_{x,y}(\mathbf{r}) = \mp \alpha \nu \frac{7\zeta(3)}{(8\pi)^2} \left(\frac{|\Delta|}{T}\right)^2\partial_{y,x}\phi(\mathbf{r}).
\end{equation}
As seen, in both regimes, the spin polarization and the direction of the current
(which is proportional to the phase gradient ${\bm
  \nabla}\phi(\mathbf{r}) $) are orthogonal to each other.

We note that similar direct magnetoelectric effect was also predicted
for superconductors and superconducting heterostructures with
intrinsic spin orbit interactions
\cite{edelstein95,edelstein05,malshukov08}. In these systems, a Rashba
spin orbit interaction is considered to be present and a generic
relation to the spin polarization obtained $S_y=\chi
(j_x/ev_F)$. Nonetheless, the coefficient $\chi$ in realistic systems
of this type
is practically negligible and proportional to $\sim \Delta_{so}/\varepsilon_F$, where
$\Delta_{so}$ is the splitting of the conducting bands due to the spin
orbit interaction and $\varepsilon_F$ is the Fermi energy. 

This issue however can be resolved in the surface channels of a 3D TI. The
surface electrons are fully spin-momentum locked and therefore they
have only one conduction band. Hence, the relation between the spin
polarization and supercurrent contains no reducing coefficient of order
of $\Delta_{so}/\varepsilon_F$ and the effect should be much stronger than
that of the conventional spin orbit coupled materials with
$\Delta_{so}/\varepsilon_F \ll 1$ discussed so far in the literature. 

\section{inverse magnetoelectric effect}

\label{inverse}
In this section, we first show that a nonequilibrium quasiparticle
injection on the surface of 3D TI induces a Zeeman-like field which is
proportional to the voltage difference
between an injector electrode and superconducting terminals that
drives the quasiparticle flow. We then calculate the dc
Josephson current through a disordered topological insulator junction
where a flow of nonequilibrium quasiparticle current is injected into
the TI region. We discuss the appearance of $\phi_0$ states which
is controlable by means of the quasiparticle flow injection.

\subsection{Zeeman-like field induced by quasiparticle injection}

In this subsection, we demonstrate that an electric current flowing
through the surface states of the 3D TI can induce a Zeeman-like
term. 
%Up to the zero order with respect to the TI/S interface transparency it is enough to consider a current flowing through a normal state of the TI, without proximity induced superconductivity. 
%
We consider a highly simple model which is sufficient to reveal this
effect. In order to more simplify our discussions, we consider fully ballistic
surface states without any disorder. In this case, the Hamiltonian is
given by Eq.~(\ref{h}) where $V_{imp}(\bm r)=\bm h(\bm r)=0$. %The
Since we have employed the quasiclassical approximation where $\mu\gg
\varepsilon$, we only consider the conduction band and assume in Eq. \ref{Green_Magnetoelectric} that
\begin{eqnarray}
{\hat{G}}(i\omega_n, \bm p)=\frac{1+ \left[ \bm n_F \times {\bm e}_z \right]{\hat{\bm \sigma}}}{2(i\omega_n+\mu-\alpha p)}
\label{G_normal_1}.
\end{eqnarray}
Moreover, we include the electron-electron interaction at the
Hartree-Fock level. Thus, the self-energy reads:
\begin{eqnarray}
{\hat{\Sigma}}= {\hat{\Sigma}^{(1)}}+{\hat{\Sigma}}^{(2)}
\label{self_general},
\end{eqnarray}
where
\begin{subequations}
\begin{eqnarray}
{\hat{\Sigma}^{(1)}}= {\rm Tr}_2 T \sum \limits_n \int \frac{d^2 p}{(2 \pi )^2} V(0,0){\hat{G}}(i\omega_n, \bm p)
\label{self_1},
\end{eqnarray}
\begin{eqnarray}
{\hat{\Sigma}}^{(2)}= - T \sum \limits_n \int \frac{d^2 p_1}{(2 \pi )^2} V(\bm p,\bm p_1){\hat{G}}(i\omega_n, \bm p_1)
\label{self_2}.
\end{eqnarray}
\end{subequations}
We also assume that $\mu$ contains the spin independent part of self-energy and focus on the spin part of ${\hat{\Sigma}= \bm \Sigma \hat{ \bm \sigma}}$. The Green function with interactions reads:
\begin{eqnarray}
{\hat{G}}(i\omega_n, \bm p )=\Big[ i\omega_n + \mu -\alpha \hat{\bm \sigma} \left[ \bm p \times {\bm e}_z \right] - \bm \Sigma \hat{\bm \sigma} \Big]^{-1}= \nonumber \\
\frac{1+(\alpha \hat{\bm \sigma} \left[ \bm p \times {\bm e}_z \right] + \bm \Sigma \hat{\bm \sigma})/\varepsilon_0}{2(i\omega_n+\mu-\varepsilon_0)}
\label{G_int},~~~~~~~~~~~~
\end{eqnarray}
where $\bm \Sigma = (\Sigma_x, \Sigma_y, \Sigma_z)$ and
\begin{eqnarray}
\varepsilon_0=\sqrt{(\alpha p_y+\Sigma_x)^2+(-\alpha p_x+\Sigma_y)^2+\Sigma_z^2}
\label{varepsilon_0}.
\end{eqnarray} 
By now substituting the Green function Eq.~(\ref{G_int}) into
Eqs.~(\ref{self_1}) and (\ref{self_2}) and assuming that $V(\bm p, \bm
p_1)=\lambda$ we find that the spin dependent part of the self-energy is:
\begin{eqnarray}
\bm \Sigma =\frac{\lambda}{4}\int \frac{d \bm p}{(2 \pi)^2}\frac{\alpha \left[\bm p \times {\bm e}_z  \right] +\bm \Sigma }{\varepsilon_0}\tanh \frac{\varepsilon_0-\mu}{2T} 
\label{self_3}.
\end{eqnarray}
In equilibrium at $V=0$, the spin dependent part of the
self-energy vanishes due to the integration over momentum. However, in
the presence of bias voltage the integral results in a nonvanishing
finite value. If we assume an electric current in the $x$ direction of
a ballistic system, we then are able to substitute $\mu \to \mu + (V/2){\rm sgn}p_x$ in Eq.~(\ref{self_3}). In the linear approximation with respect to the dimensionless interaction constant $\lambda \nu$ together with $|\Sigma| \ll \mu$ we find:
\begin{eqnarray}
\bm \Sigma =-\frac{\lambda}{4}\int \frac{d \bm p}{(2 \pi)^2}V {\rm sgn}p_x \delta(\alpha p-\mu) \left[\bm n \times {\bm e}_z  \right] 
\label{self_4},
\end{eqnarray}
where we set $T=0$ and expand Eq.~(\ref{self_3}) around $V$ to reach
at a liner response theory. Hence, the only nonzero component is $\Sigma_y$:
\begin{eqnarray}
\Sigma_y =\frac{\lambda \nu V}{2 \pi} 
\label{self_final}.
\end{eqnarray}      

In conclusion, based on a simplified model, we have shown that an
electric current can generate spin dependent exchange that results in an
effective Zeeman-like term. This induced exchange field can be controlled by the applied voltage
to inject quasiparticles. A more detailed study of the particular
connection between the induced Zeeman-like term and the injected
quasiparticles' current is an interesting problem, however, is beyond
the scope of the present paper and can be addressed elsewhere. 

\subsection{Josephson current through a 3D TI under quasiparticle
  flow injection}

The flow injection of quasiparticles into the junction generates a
nonequilibrium situation in the TI region. The Josephson current in
this nonequilibrium situation is determined not only by the condensate
wave functions in the TI region but also by the nonequilibrium
distribution function in this segment. The current can be calculated
by Eq.~(\ref{current_Usadel}). It is convenient to calculate it at the
interfaces of TI-S by exploiting the boundary conditions Eq.~(\ref{boundary_condition}):
\begin{equation}
j_x =\pm \frac{\sigma_N}{8e}{\rm Tr}_2 \int d \varepsilon \frac{1}{2\gamma}\hat \tau_z \left[  \hat g_s, \hat g_{SC} \right]^K 
\label{current_interface},
\end{equation}
where $\pm$ signs refer to the boundary conditions at $x=\pm d/2$.
The Keldysh component of the full Green function can be expressed via
the retarded and advanced components with the distribution function:
\begin{equation}
\hat g_s^K=\hat g_s^R \hat \varphi - \hat \varphi \hat g_s^A = 
\left(
\begin{array}{cc}
g_s^R \varphi - \varphi g_s^A &  f_s^R {\tilde \varphi} - \varphi  f_s^A \\
 {\tilde f_s}^R  \varphi - {\tilde \varphi}  {\tilde f_s}^A &  {\tilde g_s}^R  {\tilde \varphi} -  {\tilde \varphi}  {\tilde g_s}^A \\
\end{array}
\right)
\label{GF_distribution},
\end{equation}
where the distribution function is a diagonal matrix in the particle-hole space: 
\begin{equation}
\hat \varphi=
\left(
\begin{array}{cc}
\varphi  & 0 \\
0 & {\tilde \varphi} \\
\end{array}
\right)
\label{distribution}.
\end{equation}
There is a general relation (due to the particle-hole symmetry) between $\varphi$ and $\tilde \varphi$. For our case of spinless fermions it reduces to $\tilde \varphi(\varepsilon)=-\varphi(-\varepsilon)$.

Utilizing the distribution function, the current Eq.~(\ref{current_interface}) can be expressed by:
\begin{widetext}
\begin{eqnarray}
j_x=\pm \frac{\sigma_N}{8 e \gamma} \int d \varepsilon \left\{ -2 \left[ {\rm Re} g_{SC}^R \right](\varphi-\tilde \varphi)-
\left[ {\rm Re} f_{SC}^R \right]\tanh \frac{\varepsilon}{2T} (\tilde f_s^R e^{\pm i\phi/2}+f_s^R e^{\mp i\phi/2}+ f_s^A e^{\mp i\phi/2}+\tilde f_s^A e^{\pm i\phi/2})+ \right. \nonumber \\
\left.\left[ i{\rm Im} f_{SC}^R \right]\left( (e^{\pm i\phi/2}\tilde f_s^A-e^{\mp i\phi/2}f_s^R)\tilde \varphi+(e^{\mp i\phi/2} f_s^A -e^{\pm i\phi/2}\tilde f_s^R)\varphi \right)\right\}
\label{current_distrib}.~~~~~~~~~~~~~~
\end{eqnarray}
\end{widetext} 
Here, we assume that the left and right superconducting terminals possess
$\mp \phi/2$ macroscopic phases so that the phase difference between
the superconductors is $\phi$. The anomalous components of Green
function at the interfaces are denoted by $f_s^{R,A}$ and $\tilde f_s^{R,A}$.
In order to find the Josephson current to the leading order with
respect to the interface conductance $\gamma^{-1}$ we expand the Green
function around the bulk solution in the TI region. To obtain
Eq.~(\ref{current_distrib}) we have already taken into account this
regime and set $g^{R,A}=\pm 1$ to the first order in $\gamma^{-1}$.

\begin{figure}[!tbh]
  \centerline{\includegraphics[clip=true,width=2.5in]{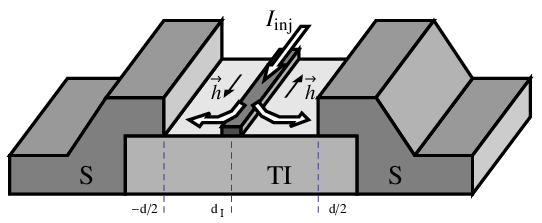}}
   \caption{Schematic of the topological insulator Josephson junction
     proposed to generate a $\phi_0$-junction by means of quasiparticle
     flow injection. The junction resides in the $xy$ plane (the same
     as Fig. \ref{system}(a)) and the injector electrode is located at $x=d_I$
     while the superconducting interfaces are assumed at $x=\pm
     d/2$. The quasiparticle flow is controlled by a voltage
     difference between the superconducting terminals and injector electrode.}   
\label{sketch}
\end{figure}

The experimental setup we propose with the electronic current injection is
schematically shown in Fig.~\ref{sketch}. The setup is made of two
superconductor terminals connected via the surface channels of a 3D TI
at $x=\pm d/2$. The junction plane resides in the $xy$ plane and a
normal electrode that injects quasiparticles into the junction is
attached at $x=d_I$. The quasiparticles flow can be induced by applying a
voltage difference between the superconducting terminals and normal
injector electrode. We assume that the voltage difference between the
two superconducting terminals is zero so that we are able to
eliminate the ambiguities and complications concerning nonequilibrium
phenomena that arise when the two superconductors possess a finite
voltage difference. Therefore, the quasiparticles current has opposite
directions within each region of the TI surface separated by the
injector electrode at $x=d_I$. The direction of quasiparticles current is shown by
white arrows in  Fig.~\ref{sketch} and we assume that the current is
uniform along the $y$ direction parallel to the S-TI interfaces.
As discussed in passing, the electric current within TI in the
$x(y)$ direction generates a finite spin polarization along the
$y(x)$ direction. Therefore, as it was shown in the previous
subsection, this electric current injection induces an effective
Zeeman-like energy $h \propto V$ in the TI region where $V$ is the voltage
difference applied between the normal injector electrode and the
superconductors. The resulting exchange fields ${\vec h}$ are shown in
Fig.~\ref{sketch}. The proposed setup generates two domains with opposite orientations for the
effective Zeeman-like term which are locked
to the direction of injected electric currents. This effective
Zeeman-like domain
scenario can be described by:
\begin{equation}
h_y(x)=\left\{
\begin{array}{rcl}
-h & , & -d/2<x<d_I \\
+h & , & d_I<x<+d/2
\end{array}
\right. .
\label{exchange_field}
\end{equation}   
To find the anomalous Green function within the TI region, we solve
the linearized Usadel equation in the surface channels:
\begin{subequations}
\begin{eqnarray}
D\kappa(\partial_x-\frac{2i}{\alpha}h_y(x))^2 f^{R,A}_s=-2i\varepsilon f^{R,A}_s, \\
D\kappa(\partial_x+\frac{2i}{\alpha}h_y(x))^2 \tilde f^{R,A}_s=-2i\varepsilon \tilde f^{R,A}_s
\label{usadel_linearized},
\end{eqnarray}
\end{subequations}
where $\kappa=\pm 1$ refer to the retarded and advanced components.

The Usadel equation should be solved together with the boundary
conditions Eq.~(\ref{boundary_condition}) at S-TI interfaces:
\begin{subequations}
\begin{eqnarray}
(\partial_x-\frac{2i}{\alpha}h_y)f^{R,A}_s|_{x= \mp d/2}=\mp
\frac{1}{\gamma} f_{SC}^{R,A}e^{\mp i \phi/2},\\
(\partial_x+\frac{2i}{\alpha}h_y)\tilde f^{R,A}_s|_{x= \mp d/2}=\pm \frac{1}{\gamma} f_{SC}^{R,A}e^{\pm i \phi/2}
\label{bc_f_lr},
\end{eqnarray}
\end{subequations}
and at the interface between the domains $(x=d_I)$:
\begin{subequations}
\begin{eqnarray}
&f^{R,A}_s|_{x=d_I -\epsilon}=f^{R,A}_s|_{x=d_I +\epsilon} , \\
&(\partial_x-\frac{2i}{\alpha}h_y)f^{R,A}_s|_{x=d_I -\epsilon}=(\partial_x-\frac{2i}{\alpha}h_y)f^{R,A}_s|_{x=d_I +\epsilon},~~~~~~~
\label{bc_f_middle}
\end{eqnarray} 
\begin{eqnarray}
&\tilde f^{R,A}_s|_{x=d_I -\epsilon}=\tilde f^{R,A}_s|_{x=d_I +\epsilon} , \\
&(\partial_x+\frac{2i}{\alpha}h_y)\tilde f^{R,A}_s|_{x=d_I -\epsilon}=(\partial_x+\frac{2i}{\alpha}h_y)\tilde f^{R,A}_s|_{x=d_I +\epsilon},~~~~~~~
\label{bc_tildef_middle}
\end{eqnarray}
\end{subequations}
where $\epsilon\rightarrow 0$.
The solution to $f^{R,A}_s$ component takes the following form:
\begin{eqnarray}
&f^{R,A}_s(x)=\frac{f_{SC}^{R,A}e^{\frac{2i}{\alpha}h_y(x)(x-d_I)}}{\gamma\lambda^{R,A}\sinh[\lambda^{R,A}d]}\left\{ e^{-i\frac{\phi}{2}-i\chi_l}\times\right. \nonumber \\
&\left. \cosh[\lambda^{R,A}(x-\frac{d}{2})]+e^{i\frac{\phi}{2}-i\chi_r}\cosh[\lambda^{R,A}(x+\frac{d}{2})] \right\}
\label{f_TI},~~~~
\end{eqnarray}
in which $\lambda^{R,A}=\sqrt{-2i\kappa \varepsilon/D}$ and
$\chi_{l,r} =2h(d/2 \pm d_I)/\alpha $. To calculate the current across
the junction, one
needs also to obtain a solution to $\tilde f^{R,A}_s$. The solution to
this component however can be simply given as $\tilde f^{R,A}_s=-f^{R,A}_s$($\phi \to -\phi$, $h_y(x) \to -h_y(x)$). 
By substituting the obtained $ f^{R,A}_s$ and $\tilde f^{R,A}_s$ into the
current definition (\ref{current_distrib}), we are able to analyze the
transport properties of the system proposed. 
The current relation (\ref{current_distrib}) can be decomposed to two components
\begin{eqnarray}
j_x=j_1+j_2
\label{current_parts},
\end{eqnarray}
where 
\begin{eqnarray}
j_1=\pm \frac{\sigma_N}{8 e \gamma} \int d \varepsilon \left\{ -2 \left[ {\rm Re} g_{SC}^R \right](\varphi-\tilde \varphi) \right\}
\label{j1},
\end{eqnarray}
and $j_2$ is expressed by the remaining parts of
Eq.~(\ref{current_distrib}) that contains the anomalous Green
functions. Note that because of opposite signs at the S-TI interfaces,
the contribution of $j_1$ component to the Josephson current
vanishes. Physically, this component represents a part of the injected
quasiparticle current. Specifically, it can be seen that $j_1$ is
equal to zero in the equilibrium simply because
$\varphi(\varepsilon)=\tilde \varphi (\varepsilon) =\tanh
\varepsilon/2T$. Therefore, $j_2$ component contains the Josephson
current and we focus on this component of the current in the following.
Since the anomalous components of Green function are of the first
order in $\gamma^{-1}$, we thus only need the distribution function
in the zeroth order with respect to this parameter for calculating
$j_2$. In this regime we have
\begin{subequations}
\begin{equation}
\varphi(\varepsilon)=\tanh \frac{\varepsilon-V}{2T},
\label{distrib1}
\end{equation}
\begin{equation}
\tilde \varphi (\varepsilon)=-\varphi(-\varepsilon)=\tanh \frac{\varepsilon+V}{2T}.
\label{distrib2}
\end{equation}
\end{subequations}
These expressions imply that the main contribution to the resistance
of the system originates from the tunnel process at the TI-S interfaces.
The following combination of the distribution functions
enables us to even decompose $j_2$ component of current to two parts
$j_{2,1}$ and $j_{2,2}$:
\begin{equation}
\varphi_\pm(\varepsilon)=\varphi(\varepsilon)\pm \tilde \varphi(\varepsilon)=\tanh \frac{\varepsilon-V}{2T} \pm \tanh \frac{\varepsilon+V}{2T}
\label{varphi_pm},
\end{equation}
\begin{widetext}
\begin{subequations}
\begin{equation}
j_{2,1}=\mp\frac{\sigma_N}{2e\gamma^2}\int d \varepsilon [{\rm Im} f_{SC}^R]\frac{\varphi_-}{2}\left( \cos(\phi-\chi){\rm Im}\left[ \frac{f_{SC}^R}{\lambda^R\sinh[\lambda^R d]} \right]+{\rm Im}\left[ \frac{f_{SC}^R}{\lambda^R\tanh[\lambda^R d]} \right] \right)
\label{j21},
\end{equation}
\begin{equation}
j_{2,2}=-\frac{\sigma_N}{2e\gamma^2}\sin (\phi-\chi)\int d \varepsilon \left\{[{\rm Re} f_{SC}^R]\tanh \frac{\varepsilon}{2T}{\rm Im}\left[ \frac{f_{SC}^R}{\lambda^R\sinh[\lambda^R d]} \right]+[{\rm Im} f_{SC}^R]\frac{\varphi_+}{2}{\rm Re}\left[ \frac{f_{SC}^R}{\lambda^R\sinh[\lambda^R d]} \right] \right\}
\label{j22},
\end{equation}
\end{subequations}
\end{widetext}
in which $\chi=\chi_r-\chi_l$ and $\pm$ at the right hand side of
Eq.~(\ref{j21}) pertinent to the left and right interfaces. As seen, $j_{2,1}$ reverses its direction at the left
and right S-TI interfaces. Consequently, this component is unable to
contribute to the current between the superconductors. This is also a part
of the injected electric current and does have no effect on the
Josephson current. 
In contrast, the current component $j_{2,2}$ does not change its sign at
the superconductor interfaces, and therefore represents the
nonvanishing current flow through the junction. This term reveals an
electrically controllable $\phi_0$-junction that we discuss in detail
in the next subsection.   

\subsection{Controllable $\phi_0$-Josephson junction}

\label{phi_0}

The Josephson current flowing through the junction is expressed by
Eq.~(\ref{j22}). The ground state of the system corresponds to zero
current and it is apparent that this condition is satisfied at a nonzero phase difference between the
superconducting leads namely at: $\phi=\chi=-4d_I h/\alpha$. According
to the previous section, this anomalous phase shift is proportional to
the voltage bias $V$ between the superconducting terminals and the
additional normal injector electrode. Therefore, the $\phi_0$ ground
state of the Josephson junction can be experimentally switched on and
off by simply controlling the quasiparticle flow injection. 

\begin{figure}[b]
  \centerline{\includegraphics[clip=true,width=8.5cm,height=3.cm]{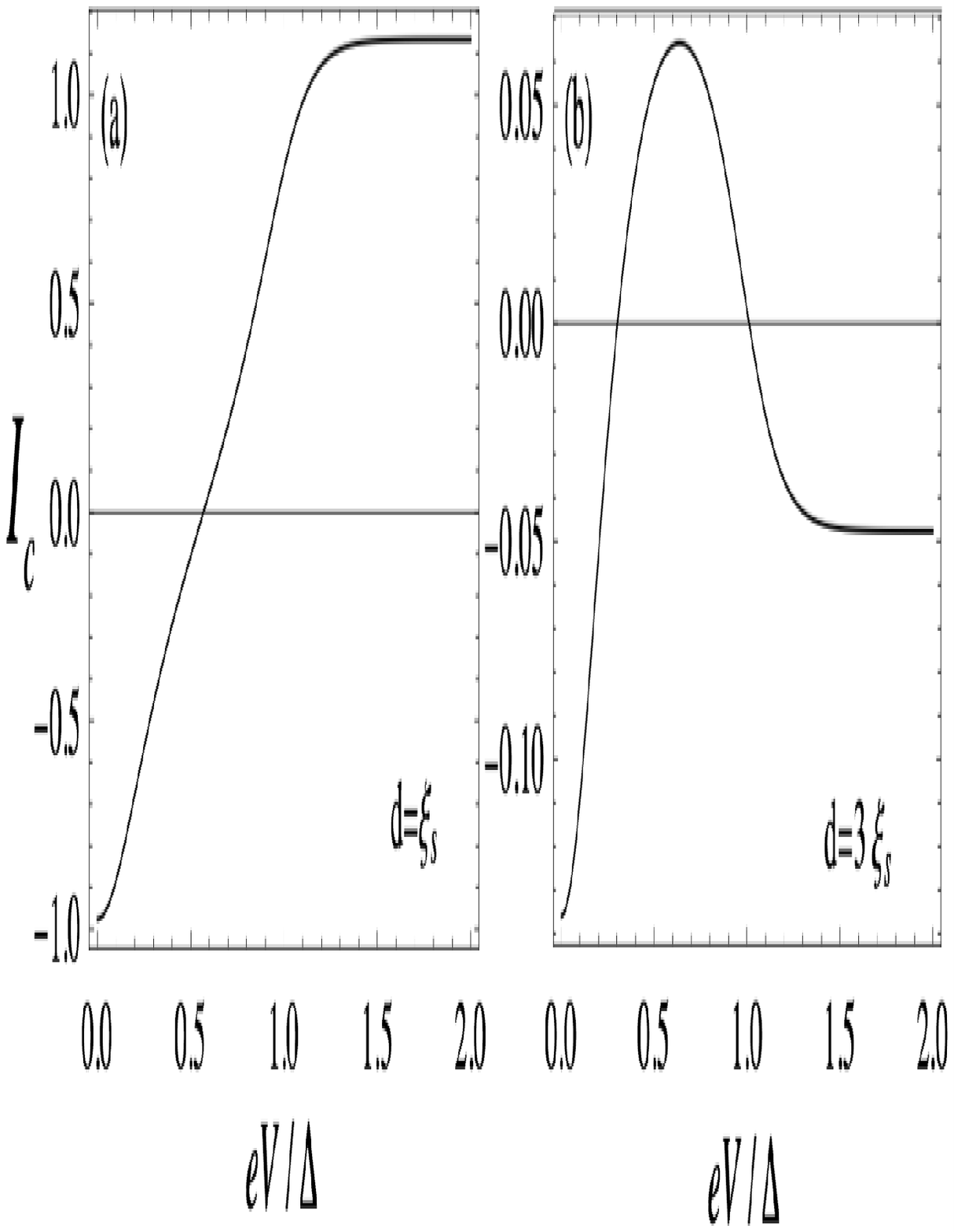}}
   \caption{Critical supercurrent as a function of voltage difference
     between the normal injector electrode and superconducting
     terminals $eV/\Delta$. In panel (a) we set the length of TI
     region equal to $d=\xi_s$
     while in panel (b) $d=3\xi_s$.}   
\label{current}
\end{figure}

The quasiparticle injection controls not only the anomalous phase
shift but also modifies the critical current of the Josephson
junction. To illustrate this fact, we plot the corresponding critical
current in Fig.~\ref{current} vs the voltage difference $V$ for two different lengths
of the TI region. The panel (a) exhibits the critical supercurrent
vs the voltage difference where the junction thickness is $d=\xi_s$
while in panel (b) we set $d=3.0\xi_s$. Here $\xi_s=\sqrt{\hbar D/|\Delta|}$ 
and this quantity is a superconducting coherence if the diffusion
constant in the superconducting terminals and TI surface are equal. The
current is measured in units of $-\sigma_N \Delta \xi_se^{-1}\gamma^{-2}$. As seen, the nonequilibrium conditions, generated by
the quasiparticles injection into the TI region, strongly alter the
critical Josephson current flowing through the junction
and can result in supercurrent reversals. We see that the critical
current in a junction of
thickness $d=\xi_s$ shows a single sign reversal while larger
thicknesses can result in multiple sign reversals. Note that
sign reversals of supercurrent due to the quasiparticles injection was
also theoretically predicted
in
conventional diffusive metals\cite{volkov95,wilhelm98} and observed in experiments 
\cite{baselmans99,huang02,crosser08}. However, the novelty of our finding is
a transition between $\chi$ and $\chi+\pi$ ground states rather than $0 \to \pi$ transition.
Our proposal offers new venues to explore more about the superconductor-topological insulator
heterostructures, identify their topological characteristics, and potentially
utilize them for technological functionalities.

\section{Conclusions}

\label{conclusions}

Motivated by recent experiments on hybrid structures of topological
insulator-supercondutor
\cite{exp_ti1,exp_ti2,exp_ti3,exp_ti4,exp_ti5,exp_ti6}, we have
generalized the quasiclassical approach using the Keldysh technique to
disordered TI-S heterostructures under nonequilibrium situations. We study the
coupling between electric charge current and spin polarization on the
surface of a three dimensional TI. Utilizing the generalized
approach, we propose a setup to experimentally achieve a controllable
$\phi_0$-Josephson junction solely by
means of a quasiparticle flow injection without resorting to an external
magnetic field or ferromagnetic element. The quasiparticles should be
injected into the TI surface from the middle of junction by applying a
voltage difference $V$ between a normal injector electrode and the
superconducting terminals. The quasiparticle currents in opposite
directions generate a Zeeman-like domain with opposite orientations. We
show that the
Zeeman-like term is directly coupled to the Josephson phase difference
and shifts the junction ground state into arbitrary values by means of
$V$ and uncover the influences of $V$ on critical supercurrent. We also discuss direct magnetoelectric effect
that appears in response to a dc Josephson current on the surface
channels and show that the induced electron spin
  polarization is much more pronounced than that of a conventional spin orbit coupled material. Our work demonstrates the great potential of topological
insulators in magnetoelectrics by superconducting hybrid configurations
due to the strong spin-momentum locking
property of TI surface channels.

\begin{acknowledgments}
A.M.B and I.V.B were supported by Grant of the Russian
Scientific Foundation No. 14-12-01290. 
\end{acknowledgments}


\begin{thebibliography}{99}
%
\bibitem{rev1} M. Z. Hasan and C. L. Kane, {Rev. Mod. Phys. {\bf 82}, 3045 (2010)}.
%
\bibitem{rev2} X.-L. Qi and S.-C. Zhang, {Rev. Mod. Phys. {\bf 83}, 1057   (2011)}.
%
\bibitem{pankratov} O. A. Pankratov, S. V. Pakhomov, and B. A. Volkov, {Solid State Comm. {\bf 61}, 93 (1987)}.
%
\bibitem{volovik} G. E. Volovik, {\it The Universe in a Helium Droplet} (Oxford University Press, Oxford, 2003).
%
\bibitem{ti1}  B. A. Bernevig, T. L. Hughes, and S-C. Zhang, {Science {\bf 314}, 1757 (2006)}.
%
\bibitem{ti2} M. Knig, S. Wiedmann, C. Brne, A. Roth, H. Buhmann,
L. W. Molenkamp, X-L. Qi, and S-C. Zhang, {Science {\bf 318}, 766 (2007)}.
%
\bibitem{ti3} D. Hsieh, D. Qian, L. Wray, Y. Xia, Y. S. Hor, R. J. Cava and M. Z. Hasan, {Nature {\bf 452}, 970 (2008)}.
%
\bibitem{ti4} Y. Xia, D. Qian, D. Hsieh, L. Wray, A. Pal1, H. Lin, A. Bansil, D. Grauer, Y. S. Hor, R. J. Cava and M. Z. Hasan, {Nat. Phys. {\bf 5}, 398 (2009)}.
%
\bibitem{ti5} H. Zhang, C-X. Liu, X-L. Qi, X. Dai, Z. Fang, and S-C.
Zhang, {Nat. Phys. {\bf 5}, 438, (2009)}.
%
\bibitem{burkov} A. A. Burkov and D. G. Hawthorn, {Phys. Rev. Lett. {\bf 105}, 066802 (2010)}.
%
\bibitem{Schwab}
P. Schwab, R. Raimondi and C. Gorini, Europhys. Lett, {\bf 93}, 67004 (2011).
%
\bibitem{zyuzin} A. A. Zyuzin, M. Alidoust, and D. Loss, {Phys. Rev. B {\bf 93}, 214502 (2016)}.
%
\bibitem{nowack} K. C. Nowack, E. M. Spanton, M. Baenninger, M. Konig, J. R. Kirtley, B. Kalisky, C. Ames, P. Leubner, C. Brune, H. Buhmann, L. W. Molenkamp, D. Goldhaber-Gordon, and K. A. Moler, {Nat. Mat. {\bf 12}, 787 (2013)}.
%
\bibitem{spanton} E. M. Spanton, K. C. Nowack, L. Du, G. Sullivan, R.-R. Du, and K. A. Moler, {Phys. Rev. Lett. {\bf 113}, 026804 (2014)}
%
\bibitem{rel1} K. Capelle and E. K. U. Gross, { Phys. Lett. A {\bf 261}, 198 (1995)}.
%
\bibitem{rel2} K. Capelle and E. K. U. Gross, {Phys. Rev. B {\bf 59}, 7140 (1999)}.
%
\bibitem{exp_ti1} I. Sochnikov, L. Maier, C. A. Watson, J. R. Kirtley, C. Gould, G. Tkachov, E. M. Hankiewicz, C. Brüne, H. Buhmann, { Phys. Rev. Lett. {\bf 114}, 066801 (2015)}.
%
\bibitem{exp_ti2} J. R. Williams, A. J. Bestwick, P. Gallagher, S. S. Hong, Y. Cui, A. S. Bleich, J. G. Analytis, I. R. Fisher, and D. Goldhaber-Gordon, {Phys. Rev. Lett. {\bf 109}, 056803 (2012)}.
%
\bibitem{exp_ti3} M. Veldhorst, M. Snelder, M. Hoek, T. Gang,	V. K. Guduru,	X. L. Wang, U. Zeitler, W. G. van der Wiel,	A. A. Golubov, H. Hilgenkamp and A. Brinkman, {Nat. Mat. {\bf 11}  417 (2012)}.
%
\bibitem{exp_ti4} J. B. Oostinga, L. Maier, P. Schüffelgen, D. Knott, C. Ames, C. Brune, G. Tkachov, H. Buhmann, and L. W. Molenkamp, {Phys. Rev. X {\bf 3}, 021007 (2013)}.
%
\bibitem{exp_ti5} S. Hart, H. Ren,	T. Wagner,	P. Leubner,
  M. Mühlbauer, C. Brüne,	H. Buhmann,	L. W. Molenkamp	and
  A. Yacoby, {Nat. Phys. {\bf 10}, 638 (2014)}.
%
\bibitem{exp_ti6} S. Lee, X. Zhang, Y. Liang, S. Fackler, J. Yong,
  X. Wang, J. Paglione, R. L. Greene, I. Takeuchi, arXiv:1604.07455.
%
\bibitem{phi0} D. B. Szombati,	S. Nadj-Perge,	D. Car,	S. R. Plissard,	E. P. A. M. Bakkers and L. P. Kouwenhoven, {Nat. Phys. {\bf 2}, 568 (2016)}.
%
\bibitem{aronov89}
A. Aronov and Y. Lyanda-Geller, JETP Lett. {\bf 50}, 431 (1989).
%
\bibitem{edelstein90}
V. Edelstein, Solid State Comm. {\bf 73}, 233 (1990).
%
\bibitem{kato04}
Y.K. Kato, R.C. Myers, A.C. Gossard, and D.D. Awschalom, Phys. Rev. Lett. {\bf 93}, 176601 (2004). 
%
\bibitem{silov04}
A.Y. Silov, P.A. Blajnov, J.H. Wolter, R. Hey, K.H. Ploog, and N.S. Averkiev, Applied Phys. Lett. {\bf 85}, 5929 (2004).
%
\bibitem{essin09}
A. M. Essin, J. E. Moore, and D. Vanderbilt, Phys. Rev. Lett. {\bf 102}, 146805 (2009).
%
\bibitem{edelstein95}
V.M. Edelstein, Phys. Rev. Lett. {\bf 75}, 2004 (1995).
%
\bibitem{edelstein05}
V.M. Edelstein, Phys. Rev. B {\bf 72}, 172501 (2005).
%
\bibitem{malshukov08}
A.G. Mal'shukov and C.S. Chu, Phys. Rev. B {\bf 78}, 104503 (2008).
%
\bibitem{krive04}
I.V. Krive, L.Y. Gorelik, R.I. Shekhter, and M. Jonson, Phys. Nizk. Temp. {\bf 30}, 535 (2004).
%
\bibitem{braude07}
V. Braude and Yu.V. Nazarov, Phys. Rev. Lett. {\bf 98}, 077003 (2007).
%
\bibitem{reinoso08}
A.A. Reynoso, G. Usaj, C.A. Balseiro, D. Feinberg, and M. Avignon, Phys. Rev. Lett. {\bf 101}, 107001 (2008).
%
\bibitem{buzdin08}
A.I. Buzdin, Phys. Rev. Lett. {\bf 101}, 107005 (2008).
%
\bibitem{kuzmanovski16}
D. Kuzmanovski, J. Linder, A. Black-Schaffer, arXiv:1605.03197.
%
\bibitem{konschelle15} 
F. Konschelle, I.V. Tokatly, and F.S. Bergeret, Phys. Rev. B {\bf 92}, 125443 (2015).
%
\bibitem{bobkova11}
I.V. Bobkova and A.M. Bobkov, Phys. Rev. B {\bf 84}, 140508 (2011).
%
\bibitem{bobkova15}
I.V. Bobkova and A.M. Bobkov, JETP Lett. {\bf 101}, 407 (2015).
%
\bibitem{ouassou16}
J. A. Ouassou, A. Di Bernardo, J. W. A. Robinson, J. Linder, arXiv:1601.07176.
%
\bibitem{heikkila00}
T.T. Heikkil${\rm \ddot a}$, F.K. Wilhelm, and G. Sch${\rm \ddot o}$n, 
Europhys. Lett. {\bf 51}, 434 (2000).
%
\bibitem{bobkova12}
I.V. Bobkova and A.M. Bobkov, Phys. Rev. Lett. {\bf 108}, 197002 (2012).
%
\bibitem{volkov95}
A.F. Volkov, Phys. Rev. Lett. {\bf 74}, 4730 (1995).
%
\bibitem{wilhelm98}
F.K. Wilhelm, G. Sch${\rm \ddot o}$n, and A.D. Zaikin, Phys. Rev. Lett. {\bf 81}, 1682 (1998).
%
\bibitem{baselmans99}
J.J.A. Baselmans, A.F. Morpurgo, B.J. van Wees, and T.M. Klapwijk, Nature (London) {\bf 397}, 43 (1999).
%
\bibitem{huang02}
J. Huang, F. Pierre, T.T. Heikkil${\rm \ddot a}$, F.K. Wilhelm, and N.O. Birge, 
Phys. Rev. B {\bf 66}, 020507 (2002).
%
\bibitem{crosser08}
M.S. Crosser, J. Huang, F. Pierre, P. Virtanen, T.T. Heikkil${\rm \ddot a}$, F.K. Wilhelm,
and N.O. Birge, Phys. Rev. B {\bf 77}, 014528 (2008).
%
\bibitem{bobkova10}
I.V. Bobkova and A.M. Bobkov, Phys. Rev. B {\bf 82}, 024515 (2010).
% 
\bibitem{ald15} M. Alidoust and K. Halterman,
New J. Phys. {\bf 17}, 033001 (2015).
%   
\bibitem{keldysh} L.V. Keldysh, Sov. Phys. JETP {\bf 20}, 1018 (1965). 
%
\bibitem{mahan} G.D. Mahan, {\it in Many-Particle Physics}, (Plenum Press, 1990).
%
\bibitem{hugdal16}
H.G. Hugdal, J. Linder, S.H. Jacobsen, arXiv:1606.01249.
%
\bibitem{usadel} K.D. Usadel, {Phys. Rev. Lett. {\bf 25}, 507 (1977)}.
%
\bibitem{bc1} A. V. Zaitsev, Sov. Phys. JETP 59, 1015 (1984).
%
\bibitem{bc2} M. Y. Kuprianov and V. F. Lukichev, Sov. Phys. JETP 67,
  1163 (1988).


\end{thebibliography}
\end{document}